\documentstyle[12pt,epsf,epsfig]{article}

\begin{document}

\begin{titlepage}

\vspace*{0.1cm}   
\begin{flushright}  
DTP/96/42  \\
May  1996  \\   
\end{flushright}   
\vskip   1.2cm   

\begin{center}
{\Large\bf   
A leptophobic massive vector boson \\[2mm]  
at the Tevatron and the LHC } 
\vskip 1.cm 
{\large  M.~Heyssler\footnote{email: M.M.Heyssler@durham.ac.uk}}  
\vskip .3cm
{\it  Department  of Physics,  University of Durham, \\ 
Durham DH1 3LE, England }\\

\vskip 1cm

\end{center}

\begin{abstract}

Recent  measurements of the single  inclusive jet cross section at the
Tevatron by the CDF Collaboration  maybe suggest a modified picture of
QCD  in the  large  $E_T$  range.  One  possible  explanation  of  the
measured  jet excess is the  introduction  of a neutral  heavy  vector
boson  $Z'$.  A  parameter  fit of this new model to the CDF  data, in
leading order  perturbation  theory, is performed, and the question of
how the  corresponding  single  inclusive  jet cross  sections and the
dijet angular distributions at the LHC are affected by this additional
$Z'$ is discussed.  We conclude that the $Z'$ will play a pivotal role
for typical LHC  centre--of--mass  energies,  thus  providing a direct
test of this theory.

\vskip 2cm


\end{abstract}

\vfill 

\end{titlepage}

\newpage

\section{Introduction}
\label{intro.ms}

Recent  data from the CDF  Collaboration  \cite{CDF93}  on the  single
inclusive  jet cross  section  at the  Tevatron  indicate  a  possible
disagreement  with QCD at high  transverse jet energies.  The reported
excess rate exceeds NLO QCD  calculations  by 10--50\% for  $200$~GeV$
<E_T<400$~GeV.  One has to be cautious in drawing rash conclusions for
the  evidence of new physics, as the D0  Collaboration  have  reported
{\em  agreement}  with  QCD in the  same  measured  jet  energy  range
\cite{D096}.  Still the systematic  errors in both experiments are too
large to enable  definite  conclusions  to be drawn.  But also the SLC
and especially the LEP Collaborations  \cite{LEP95}  ALEPH, DELPHI, L3
and OPAL reported deviations measured in high precision experiments on
the  ratios  $R_{b,c}  =  \Gamma(Z  \rightarrow  b\bar{b},  c\bar{c})/
\Gamma(Z  \rightarrow  hadrons)$.  Compared to the  predictions of the
Standard  Model, they find a too {\em large}  value for $R_b$ at about
the  $3.5\sigma$  level and a too {\em small} value for $R_c$ at about
the  $2.5\sigma$  level.  As $R_b$ and $R_c$ are correlated  one might
e.g.  arbitrarily  set $R_c$ to the LEP1  experimental  value, but the
excess of $R_b$, now on a $3.0\sigma$ level, remains.

Discussions  continue on how to understand  the CDF and (or)  LEP1/SLC
data from a  phenomenological  point of view if the disagreement  with
the Standard Model  predictions is taken  literally.  For the CDF data
there are efforts to explain the observed effects in terms of modified
parton  distributions  \cite{CDFMODELS},  quark  substructures,  quark
resonances or some more exotic models  \cite{Bander96}.  Independently
the measured $R_{b,c}$ values were treated in the framework of various
extensions  to the Standard  Model  \cite{LEPMODELS}.  However, in two
recent publications by Altarelli {\em et al.}  \cite{Altarelli96}  and
Chiapetta {\em et al.}  \cite{Chiapetta96}  both CDF and LEP1/SLC data
are treated on the same level and are described by a universal effect:
the  introduction of an additional  very massive  neutral vector boson
coupled to the neutral quark sector of the Standard  Model.  This $Z'$
boson has the  feature  that it  couples  very  strongly  to $u$-- and
$d$--type quarks and contributes to the standard boson $Z$ decay via a
weak $Z'$--$Z$ mixing angle $\xi$.  Analysing the experimental data of
the CDF and LEP1/SLC  Collaborations  allows a global parameter fit of
the $Z'$  model,  and it was shown in  Ref.~\cite{Altarelli96}  that a
best--fit set of  parameters  can be found, to explain  simultaneously
the CDF jet data and the measured $R_{b,c}$ values.

We shall exploit this idea and undertake a global analysis of the $Z'$
model in the  context  of the CDF data  only, to show the  differences
with the results of  Refs.~\cite{Altarelli96,Chiapetta96}  if one only
takes the CDF data into account.  But the main intention of this paper
is to present  predictions of the $Z'$ model for further  measurements
at the Tevatron,  like dijet angular  distributions,  and of course at
the LHC  $pp$--collider.  As the  $Z'$  model  seems a  quantitatively
plausible  description  of  the  observed  deviations  so  far,  it is
important to give predictions for future experiments to either support
or discard this explanation.

To give a brief  outline of this  paper, we discuss  the $Z'$ model in
Section~\ref{sec2.ms}     and    introduce    its     parametrisation.
Section~\ref{sec3.ms}  focuses on the present  and future  data at the
Tevatron.  We fit the $Z'$ model  parameters  to the CDF jet data and,
as a  first  application,  give  predictions  for  the  dijet  angular
distributions  in LO QCD  with  the  $Z'$  contribution  included.  In
Section~\ref{sec4.ms}  we apply our {\em  best--fit} $Z'$ model to the
LHC.  Again we calculate the single  inclusive jet cross  sections and
the  dijet  angular   distributions.  Finally,   Section~\ref{sec5.ms}
summarises  our results,  underlines the most  important  features and
discusses open problems.


\section{The $Z'$ model}
\label{sec2.ms}

The   $Z'$   model   introduced   by   Altarelli   {\em  et   al.}  in
Ref.~\cite{Altarelli96}  and  independently  by Chiapetta {\em et al.}
in  Ref.~\cite{Chiapetta96}  to explain recent experimental deviations
from  the  Standard   Model,  has  the  remarkable   feature  (as  the
experimental  data demand) that the axial and vector  couplings of the
$Z'$,  especially to $u$--type  quarks, are quite large.  It will turn
out that the  effective  $Z'u\bar{u}$  coupling is of the order of the
strong  (QCD)  coupling  constant  $\alpha_S$.  Especially  for  large
energies  (transverse jet energies $E_T$) the contributions due to the
additional $Z'$ are becoming dominant and for a fitted set of coupling
parameters will for example cure the measured jet excess.  We shall be
very cursory in the presentation of the $Z'$ model as it is treated in
almost complete analogy to the $Z$ boson of the Standard Model and has
been          already          broadly           discussed          in
\cite{Altarelli96,Chiapetta96,LHC90}.

To introduce the $Z'$, the neutral  sector of the Standard  Model with
the underlying ${\rm SU}(3)_{\rm C}\times{\rm SU}(2)_{\rm L}\times{\rm
U}(1)_{\rm  Y}$ gauge group is extended by an  additional  term in the
neutral--current Lagrangian
\begin{eqnarray}
\label{nclagrangian}
{\cal{L}}'_{\rm NC} &=& \frac{g}{\cos(\Theta_W)} {J_\mu'}^0 {Z'}^\mu
\nonumber \\
&=& \frac{g}{2\cos(\Theta_W)} \sum\limits_f 
\overline{\Psi}_f\gamma_\mu
\left( v'_f + a'_f\gamma^5 \right) \Psi_f {Z'}^\mu. 
\end{eqnarray}
The neutral current ${J_\mu'}^0$  includes the axial $a_f'$ and vector
$v'_f$  coupling  strengths of the $Z'$.  In the Standard  Model there
are  three   free   coupling   parameters   for  the  $Z$  boson:  the
left--handed  coupling to the  $(u,d)_{\rm  L}$  doublets  and the two
right--handed  couplings  $u_{\rm  R}$ and  $d_{\rm  R}$.  To preserve
these  degrees of  freedom,  we follow  the quark  family--independent
parametrisation  for $u$-- and $d$--type quarks in  \cite{Altarelli96}
for $a_f'$ and $v'_f$
\begin{eqnarray}
\label{coupling}
v'_u &=& x + y_u, \quad   a'_u = -x + y_u,\nonumber \\
v'_d &=& x + y_d, \quad\, a'_d = -x + y_d.
\end{eqnarray}
All  couplings  to  leptons  are  set  to  zero  (leptophobic   $Z'$):
$v'_l=v'_\nu=0$  and  $a'_l=a'_\nu=0$.  In   \cite{Altarelli96}   this
constraint  was due to the fact that only  deviations  from  $R_b$ and
$R_c$ have been reported by the LEP1/SLC measurements.  Apart from $x,
y_u$ and  $y_d$  there are two more  parameters  included  in the $Z'$
model:  the mixing  angle  $\xi$  between  $Z$ and $Z'$ as well as the
mass $M_{Z'}$ of the $Z'$.  With these  parameters we can   also fully
determine the total decay width of the $Z'$
\begin{equation}
\label{width}
\Gamma_{Z'} = \frac{G_FM_Z^2}{2\sqrt{2}\pi} N_c M_{Z'}
\left( {v_f'}^2 + {a_f'}^2 \right),
\end{equation}
where $N_c$ is the number of quark colours and $G_F$ denotes the Fermi
constant.

From fitting various electroweak  observables to the LEP1/SLC data and
taking   the   CDF   results    into    account,    the   authors   of
Ref.~\cite{Altarelli96}  find  as  best  set of  parameters:  $x=-1.0,
y_u=2.2, y_d=0.0$ and  $\xi=3.8\cdot10^{-3}$  with the $Z'$ mass fixed
in this analysis to be $M_{Z'}=1$~TeV.  This parameter space gives the
best  numerical   compromise  to   simultaneously   obtain  acceptable
coincidence  with the values for $R_{b,c}$  {\em and} the measured CDF
jet rate.  Such a heavy vector boson is in  accordance  with the lower
mass  limit of 412~GeV  (at a 95\%  confidence  level)  reported  from
$p\bar{p}$--collider  experiments in a search for a new neutral vector
boson (with standard  couplings)  \cite{PDG94}.  The dependence on the
$y_d$ parameter was found to be weak \cite{Altarelli96}, such that the
somewhat  arbitrarily  choice of  $y_d=0.0$  was used as an input.  We
shall exploit these results and concentrate on finding the best set of
parameters for $x$ and $y_u$  describing  the CDF data within the $Z'$
model,  with  $\xi,  y_d$  and  $M_{Z'}$  fixed  to the  values  given
above\footnote{ As we restrict ourselves to fitting the CDF data only,
the mixing angle $\xi$ does not appear as a free  parameter.  However,
because  we  later  want  to  calculate  $R_{b,c}$  for  the  sake  of
comparison with the Standard Model  predictions  and the LEP1 data, we
shall  fix  $\xi$  to the  value  given  by  Altarelli  {\em  et  al.}
\cite{Altarelli96}.}.


\section{Fit to the CDF single inclusive jet data}
\label{sec3.ms}

In this  section we shall  perform a global  $\chi^2$  fit of the $Z'$
model parameters $x$ and $y_u$ discussed in  Section~\ref{sec2.ms}  to
the 1992--93 measurements of the single inclusive jet cross section by
the CDF Collaboration \cite{CDF93}.

In leading order (LO) QCD the process  $AB\rightarrow  jet + X$ can be
parametrised by \cite{Sterman95}
\begin{eqnarray}
\label{single}
\frac{d^2\sigma}{dE_Td\eta}(AB &\rightarrow& jet + X) = 
4\pi\alpha_S^2(Q^2) 
\frac{E_T}{s^2} \\ \nonumber
&\times&\sum\limits_{abcd}\int\limits_{x_a^{\rm min}}^{1} 
\frac{dx_a}{2x_a-x_Te^{\eta}} 
\frac{f_{a/A}(x_a,Q^2)}{x_a} \frac{f_{b/B}(x_b,Q^2)}{x_b} 
|\overline{\cal{M}}_{ab\rightarrow cd}|^2,
\end{eqnarray}
in terms of the  transverse  energy $E_T$ of the  observed jet and the
directly  measured  pseudorapidity  $\eta$.  The  expressions  for the
squared and averaged matrix elements of the subprocesses  contributing
to  $|\overline{\cal{M}}_{ab\rightarrow  cd}|^2$  in  LO  due  to  the
partons  $a,b,c$ and $d$ being  quarks,  antiquarks  or gluons, can be
found in e.g.  \cite{Combridge77}  or any standard  QCD  textbook.  We
integrate  over the  kinematical  variable  $x_a$  only,  with  $x_b =
x_ax_Te^{-\eta} /  (2x_a-x_Te^\eta)$  and $x_a^{\rm min} = x_Te^\eta /
(2-x_Te^{-\eta})$.  The variable  $x_T$ is the scaled  counterpart  of
$E_T$ being $x_T = 2E_T/\sqrt{s}$.  Eq.~(\ref{single}) fully describes
the single inclusive jet cross section.  For the parton  distributions
$f_{(a,b)/(A,B)}(x_{(a,b)},Q^2)$  we use the MRS(A$'$)  set of partons
described in Ref.~\cite{Martin95}.

The inclusion of the $Z'$ into the formalism is  straightforward.  One
has to  calculate  those  matrix  elements in which the  incoming  and
outgoing partons are quark and antiquark pairs.  The only  constraints
at  the   $Z'q\bar{q}$   vertices  are   colour--charge   and  flavour
neutrality.  All   possible   $Z'$   exchanges   in  the   $s$--   and
$t$--channels have to be taken into account (cf.  Fig.~\ref{kin.fig}).
The analytic expressions for these amplitudes are for example cited in
\cite{Altarelli96,LHC90} and will not be repeated here.

Throughout this work we shall restrict ourselves to the LO calculation
of the jet cross  sections.  For small  values of $|\eta|$ it has been
shown in e.g.  \cite{rscale}  that for single inclusive jet production
at high transverse energies the next--to--leading  order (NLO) and the
LO  calculations  only  differ by a constant  factor,  independent  of
$E_T$, if one chooses  $\mu=E_T/2$ as the  underlying  renormalisation
scale.  This  renormalisation  scale is imbedded into our calculations
in the form of the four--momentum transfer $Q^2=\mu^2$ as the defining
scale for the running coupling constant $\alpha_S(Q^2)$ and the parton
distributions.  The difference  between LO and NLO is then reported to
be less  than 10\% and  independent  of $E_T$  for  $E_T>$100--200~GeV
\cite{rscale}.  The lower bound on $E_T$  depends on the set of parton
distributions used and the value of $\Lambda_{\rm  QCD}$  implemented.
For  the  MRS(A$'$)  set  the  QCD  scale  parameter  is  found  to be
$\Lambda_{\overline{\rm  MS}}^{(N_f=4)} = 231$~MeV, which  corresponds
to  $\alpha_S(M_Z^2)  =  0.113$  \cite{Martin95}.  The  MRS(A$'$)  NLO
calculation  was shown to be in good agreement  \cite{CDF93}  with the
CDF single inclusive jet data up to $E_T\simeq 200$~GeV.  We therefore
normalise  our LO  calculations  of the  single  inclusive  jet  cross
section to the CDF  measurements in the range  $150$~GeV$<E_T<200$~GeV
as shown in  Fig.~\ref{lot.fig}.  The dashed curve  represents  the LO
QCD calculation according to Eq.~(\ref{single}), the solid curve shows
the {\em  corrected} LO  calculation  normalised to the CDF data which
are  also  presented.  For   $130$~GeV$<E_T<200$~GeV   the  difference
between  the  central  values  of the CDF data and the  normalised  LO
calculation is less than 5\%.  The normalisation factor is found to be
${\cal{N}}=0.091\pm0.003$ according to the reported statistical errors
of the CDF  data.  Comparing  our  results  with  those  presented  in
\cite{rscale} we conclude that for  $E_T>130$~GeV  and $\mu=E_T/2$ our
LO  calculation  is  adequate  to NLO  assuming  the  constant  factor
${\cal{N}}$.  For our  $\chi^2$  analysis  of the  CDF  data we  shall
therefore   use   the   normalised   LO   calculation   presented   in
Fig.~\ref{lot.fig}.

The  CDF   Collaboration   reported  a  significant   jet  excess  for
$E_T>200$~GeV  \cite{CDF93}.  In the  inset of  Fig.~\ref{lot.fig}  we
present the  conspicuous  deviations  of the CDF data in the  measured
energy range to our LO calculation  in per cent.  The solid line shows
the  anticipated  {\em best--fit}    calculation  in LO with  the $Z'$
incorporated  and  the  smallest  achievable  $\chi^2$  value.  Let us
therefore now briefly discuss our fit of the $Z'$ model parameters $x$
and $y_u$ to the CDF data.

\subsection{$\chi^2$ analysis of the $Z'$ model}
\label{sec3.1.ms}

The {\em  qualitative}  difference  of our $Z'$  model  fit to that of
Altarelli {\em et al.}  \cite{Altarelli96} is that we only concentrate
on the CDF data and  disregard  the values for the quark  ratios $R_b$
and  $R_c$  measured  at  the  LEP1/SLC   colliders  for  the  moment.
Furthermore   we  are   using  a   different   renormalisation   scale
($\mu=E_T/2$ rather than $\mu=E_T$) and therefore approach NLO results
in  a  natural  way   \cite{rscale}.  We  also   perform  an  implicit
integration   over   the   pseudorapidity    $\eta$   in   the   range
$0.1\leq|\eta|\leq0.7$, more in line with the  experimental  cuts used
by the CDF Collaboration.

Nevertheless we expect our {\em best--fit} parameters to be very close
to those found in  \cite{Altarelli96}  such that we constrain three of
the five  parameters (cf.  Section~\ref{sec2.ms})  in exact analogy to
this work, namely $\xi=3.8\cdot10^{-3}$ (mixing angle), $M_{Z'}=1$~TeV
($Z'$ mass) and  $y_d=0.0$.  We are left with two  parameters  $x$ and
$y_u$ to define the  $\chi^2$  distribution  of our  problem.  We show
$\chi^2(x,y_u)$  in   Fig.~\ref{chi.fig}a.  Note  that  the  pure  QCD
calculation yields $\chi^2(0,0)=45.14$.  Fig.~\ref{chi.fig}b shows the
95.4\%  confidence  ellipse  (2$\sigma$ for the normal  distribution).
The statistical  analysis was performed using the programming  package
of  Ref.~\cite{C91}.  While $x$ is bound according to this analysis to
a very narrow band, the parameter  $y_u$ covers a much broader  range.
The  narrowness of the $x$ range is due to the fact that it influences
both $u$-- and $d$--type  quarks  simultaneously,  and  therefore  its
variation is much more constrained.

Finally  in  Fig.~\ref{chi.fig}c  we  present  the  68.3\%  confidence
ellipse    (1$\sigma$ for the normal  distribution)  and deduce    the
best--fit parameters of our analysis to be
\begin{eqnarray}
\label{parameters}
   x   &=& -1.0, \quad y_u = 2.8, \nonumber\\
   {\rm with}\qquad y_d &=& 0.0, \quad 
   M_{Z'} = 1\;{\rm TeV}, \quad \xi=3.8\cdot10^{-3}.
\end{eqnarray} 
Altarelli {\em et al.}  \cite{Altarelli96}  report a slightly  smaller
value of $y_u=2.2$.  This is mainly due to the included  $R_{b,c}$ fit
as well as to the  differences in the analysis  procedure as discussed
above.  The  improved  result  for  the  single  inclusive  jet  cross
section, due to  incorporated  $Z'$  exchange with the  parameters  of
(\ref{parameters}),    was    already    shown   in   the   inset   of
Fig.~\ref{lot.fig}.  Note  that  with  this  set  of  parameters   the
coincidence  with the  experimental  LEP1  values  of $R_b$ and  $R_c$
\cite{LEP95}  is still  better than the  predictions  by the  Standard
Model, as shown in Table~\ref{tab}.

\begin{table}[t]
\begin{center}
\begin{tabular}{cccc}\hline\hline
\rule[-3mm]{0mm}{8mm}
      & {\rm our fit} & {\rm LEP1}      & {\rm Standard Model} 
\\ \hline
\rule[-3mm]{0mm}{8mm}
$R_b$ & 0.2194        & $0.2219\pm0.0017$ & $0.2156\pm0.0005$      
\\ 
\rule[-3mm]{0mm}{8mm}
$R_c$ & 0.1642        & $0.1543\pm0.0074$ & $0.1724\pm0.0003$      
\\ \hline\hline
\end{tabular}
\end{center}

\caption[]{\label{tab}Comparison  of the  values  $R_{b,c}$  from  our
calculation   including  the  $Z'$  model  and  the  {\em   best--fit}
parameters      of      (\ref{parameters})      with      the     LEP1
measurements~\cite{LEP95}  and the predictions of the Standard Model.}
\end{table}

With (\ref{parameters}) and $M_Z=91.18$~GeV we find a total $Z'$ decay
width according to Eq.~(\ref{width}) of $\Gamma_{Z'}=644.2$~GeV.  This
should  be  compared  to the  value  for the  standard  $Z$  boson  of
$\Gamma_Z=2.493\pm0.004$~GeV     \cite{PDG94}.    Our     value    for
$\Gamma_{Z'}$  exceeds  the  one  assumed  by  Chiapetta  {\em et al.}
\cite{Chiapetta96} by a factor of three.  From Eq.~(\ref{coupling}) we
find the vector and axial  couplings of the $Z'$ to  $u$--type  quarks
being   $v'_u=1.8$  and  $a'_u=3.8$.  These  values  should  again  be
compared  with  the  Standard  Model   predictions   \cite{PDG94}   of
$v_u=0.19$ and $a_u=0.50$ for the $Z$ boson.  As already  mentioned in
Section~\ref{sec2.ms}, the effective $Z'u\bar{u}$ coupling is of order
$({v_u'}^2+{a_u'}^2)\alpha_W \sim \alpha_S$.  So the main contribution
of the $Z'$  follows  from its  coupling to  $u$--type  quarks with an
absolute  strength that is  comparable  to QCD itself.  The effects of
this coupling can be observed in the inset of Fig.~\ref{lot.fig} where
for  $E_T\sim400$~GeV,  the $Z'$ contribution  already equals the pure
QCD contribution.

Before we shall  answer the  question  of how this $Z'$ model with the
new  parameter  fit will affect jet physics at the LHC we shall  first
discuss the  comparison  of our results to the already  available  and
future data of the dijet angular distributions at the Tevatron.

\subsection{Comparison  with  the  measurements  of  the  dijet  cross
sections at the Tevatron}
\label{sec3.2.ms}

The leading order differential dijet cross section in a hadron--hadron
collision can be expressed in terms of the centre--of--mass scattering
angle  $\cos(\Theta^\star)$  and the  invariant  mass of the two  jets
$M_{jj}$ \cite{Sterman95}
\begin{eqnarray}
\label{angular}
\frac{d\sigma}{d\cos(\Theta^\star)dM_{jj}}(AB &\rightarrow& jet_1 +
jet_2 + X) = 
4\pi\alpha_S^2(Q^2)
\frac{1}{8M_{jj}^2} \\ \nonumber
&\times& \sum\limits_{abcd}\int\limits_{x_a^{\rm min}}^1 
dx_a f_{a/A}(x_a,Q^2)f_{b/B}(x_b,Q^2)
|\overline{\cal{M}}_{ab\rightarrow cd}|^2,
\end{eqnarray}
with  $x_a^{\rm   min}=M_{jj}^2/s$   and   $x_b=M_{jj}^2/x_as$.  Again
$a,b,c$ and $d$ denote the different  types of partons and $A$ and $B$
the scattering  hadrons.  The cross section is again  factorised  into
one part that includes the information on the parton densities  inside
the hadrons and the averaged matrix element  squared part that carries
the $\cos(\Theta^\star)$ information.  So the jet angular distribution
is sensitive to the form of the $2\rightarrow 2$ matrix elements.  For
small  angles, the partonic  contributions  to the total  differential
cross     section    show    a    typical     Rutherford     behaviour
$(\sim\sin^{-4}(\Theta^\star/2))$.  To remove this  singularity  it is
convenient  to plot  the  angular  distribution  in  terms of  another
variable  $\chi$ defined  as\footnote{To  minimise  confusion we shall
always denote  the angular  variable by $\chi$ whereas the statistical
variable  is  denoted  by  $\chi^2$.}  
\begin{equation}   
\label{chi}   
\chi  = \frac{1+|\cos(\Theta^\star)|}{1-|\cos(\Theta^\star)|}.  
\end{equation}  
It is clear that  $\chi\in  \lbrack  1,\infty  \rbrack$.  In the small
angle region ($\chi$ large) one expects therefore $d\sigma/d\chi \sim$
const.  as $d\chi/d\cos(\Theta^\star) \sim \sin^{-4}(\Theta^\star/2).$

The  vindication  of  restricting  ourselves to a LO  calculation  has
already  been  discussed  in the  case  of the  single  inclusive  jet
analysis.  We  concluded  that for $E_T >  130$~GeV  LO is a very good
approximation  to NLO (cf.  Fig~2)  if one  chooses  $\mu  =E_T/2$  as
underlying  renormalisation  scale, and takes a  normalisation  factor
${\cal{N}}$  into  account.  The dijet mass,  however, is connected to
the transverse jet energy via the relation
\begin{equation}
\label{djmass}
M_{jj} = 2 E_T \cosh(|\eta^\star|),
\end{equation}
where we introduce the centre--of--mass  pseudorapidity  $\eta^\star =
(\eta_1-\eta_2)/2$    (with   $\eta_1$   and   $\eta_2$    being   the
pseudorapidities in the lab--frame).

With  $\cos(\Theta^\star)=\tanh(\eta^\star)$  and  Eq.~(\ref{chi})  we
find  that   $\chi=e^{2|\eta^\star|}$.  Therefore   Eq.~(\ref{djmass})
yields  $M_{jj}  =  E_T(\sqrt{\chi}  +  1/\sqrt{\chi})$.  So one could
expect that for large $M_{jj}$  $(M_{jj}>260$~GeV) and small values of
$\chi$  our   argumentation   concerning   the   validity  of  the  LO
approximation   might  still  hold.  However,  if  there  is  a  large
transverse  boost $\eta_{\rm  boost}=(\eta_1  +\eta_2)/2$ to the dijet
system   then   $\chi$  can  become  as  large  as   $|\eta^\star|   =
|\eta_1-\eta_{\rm  boost}|$  but LO can  still be  adequate  to NLO if
$|\eta_1|$ is small.  On the other hand $|\eta_{\rm  boost}|$ could be
small and $|\eta_1|$  large:  in this case the LO  description  fails.
So one has to be cautious with the  argumentation.  However Ellis {\em
et al.}  \cite{Ellis92}  also determined the scale $\mu$ for which the
calculation  approximately  reproduces  the less scale  dependent  NLO
result in the case of dijet production.  If we express their result in
terms of the variable $\chi$, one finds
\begin{equation}  
\label{ellis}  
\mu\approx k(\chi)\frac{E_T}{2},              
\end{equation}             
with  $k(\chi)=(\chi+1)/(\chi^{0.85}+\chi^{0.15})$.  For  $\chi=1$  we
find $\mu\approx  E_T/2$, the value for the  renormalisation  scale we
were  using  throughout.  We  conclude  that also in the case of dijet
production this scale yields a reliable approximation to NLO (at least
in  the   small   $\chi$   range).  For   $\chi=5,10,20$   one   finds
$k(\chi)=1.15,1.29,1.39$ such that nearly the complete range for small
values  of  $\chi$  is in  approximate  accordance  with  NLO for $\mu
=E_T/2$.  However, to approach the NLO result in a pure LO calculation
as good as possible, we shall use the {\em effective}  renormalisation
scale  of  Eq.~(\ref{ellis})  for  the  study  of  the  dijet  angular
distributions  throughout  this section.  With this choice of $\mu$ we
do not  have to  worry  about  the  normalisation  factor  ${\cal{N}}$
introduced for the case of the single inclusive cross section.

We show in Fig.~\ref{dj1.fig}  our calculations in lowest order QCD as
well as in the extended model  (QCD+$Z'$)  with the coupled $Z'$.  The
$Z'$  model  parameters  are  again  fixed  to  the  values  given  in
(\ref{parameters}).  We compare our  results  first with the data from
the CDF  Collaboration  of 1992  \cite{CDF92}.  They  measured the jet
angular  distribution  with a jet data  sample of  $4.2$~pb$^{-1}$  in
three different dijet mass regions (Fig.~\ref{dj1.fig}a--c).  Only the
statistical  errors are shown.  The systematic  errors are reported to
be   $5$--$10\%$    \cite{CDF92}.   The   kinematical   cut   on   the
centre--of--mass  pseudorapidity  was chosen to be  $|\eta^\star|<1.6$
for     $240$~GeV$<M_{jj}<475$~GeV     and    $M_{jj}>550~$GeV;    and
$|\eta^\star|<1.5$   for    $475$~GeV$<M_{jj}<550$~GeV.   Again   with
$\chi=e^{2|\eta^\star|}$ we get upper bounds for $\chi$, such as $\chi
< 24.5$ for  $\eta^\star  < 1.6$ and $\chi < 20.0$ for  $\eta^\star  <
1.5$.  All cross  sections in  Fig.~\ref{dj1.fig}  are  normalised  to
unity in the  corresponding  $\chi$ intervals, and integrated over the
given  $M_{jj}$  range.  As the cross  section falls very steeply in a
given $\chi$ bin ($\propto  1/M_{jj}^3$),  we introduce a cut--off for
the  dijet  mass  in   Fig.~\ref{dj1.fig}c   of  $M_{jj}=700$~GeV.  An
analysis of the cut--off dependence showed that any higher upper bound
on $M_{jj}$ changes the result by less than 2\%.

From a first look   at  Fig.~\ref{dj1.fig}  we notice that all angular
cross sections are rising for higher values of $\chi$.  This is due to
the  fact  that  we  incorporated   our  running   coupling   constant
$\alpha_S(Q^2)$  with  $Q^2=k^2(\chi)E^2_T/4$.  The  $Q^2$  scale is a
function  of  $M_{jj}$  and $\chi$.  This can be deduced by  examining
Eq.~(\ref{djmass}).    It    follows     directly    that    $Q^2    =
M_{jj}^2\chi/4(\chi^{0.85}+\chi^{0.15})^2$   with  $Q^2_{\rm   max}  =
M_{jj}^2/16$.  For  larger  values of $\chi$  the  values of $Q^2$ are
therefore becoming smaller.  The partons are probed at lower energies,
but the  effective  coupling  $\alpha_S(Q^2)$  is  rising  as $Q^2$ is
shrinking.

A second feature  becomes  transparent  from  Fig.~\ref{dj1.fig}:  the
influence of the $Z'$ is less  striking for small and  moderate  dijet
masses as shown in  Fig.~\ref{dj1.fig}  but becomes more important for
higher  values of  $M_{jj}$.  We have to recall  that a dijet  mass of
$M_{jj}=500$~GeV for $\chi=2.5$ corresponds to a transverse jet energy
$E_T=226$~GeV,  whereas a dijet mass of $M_{jj}=1000$~GeV  corresponds
to  $E_T=452$~GeV  for the  same  value  of  $\chi$.  The $Z'$  model,
however, has been  constructed  in such way that its influence is only
felt for $E_T>200~$GeV.  Therefore only calculations with a relatively
high dijet mass at  $\sqrt{s}=1.8$~TeV  are substantially  affected by
the $Z'$ boson.  But already for $\langle M_{jj}\rangle  =500$~GeV and
$\langle  M_{jj}\rangle  =600$~GeV the presence of the additional $Z'$
becomes  transparent (cf.  Fig.~\ref{dj1.fig}b,c),  especially for the
large--angle--scattering ($\chi$ small).  This is due to the fact that
such a massive vector boson acts like an effective contact interaction
\cite{Eichten89} (Fig.~\ref{kin.fig}) between the four quarks at small
energy  transfers  in the $s$-- and  $t$--channels.  As, for  example,
$|t|=M^2_{jj}/(\chi+1)$  we obtain  $|t|\ll  M^2_{Z'}$, if $\chi\gg 1$
and  ${\cal{O}}(M^2_{jj})\simeq{\cal{O}}(M^2_{Z'})$.  Because  of  the
general    form    of    the    $Z'$    matrix    elements    squared,
$|\overline{{\cal{M}}}_{Z'}|^2\propto                         1/\left(
(t-M_{Z'}^2)^2+M_{Z'}^2\Gamma_{Z'}^2\right)$ \cite{Altarelli96,LHC90},
we  find  the  $Z'$  contribution  becoming  flat  for  large  $\chi$.
Therefore the observed  enhancement of the dijet cross sections due to
this  additional  vector  boson only takes  place for small  values of
$\chi$.

The  comparison  with the CDF data  should be  regarded  only as being
illustrative,   as  for  larger  values  of  $\chi$  the  NLO  and  LO
calculations slightly differ.  The main purpose of  Fig.~\ref{dj1.fig}
is to show the influence of the $Z'$ on the pure QCD calculations.  As
we  expected  from the {\em a priori}  construction  of the $Z'$,  its
presence  is  emphatically  felt  for  higher  dijet  masses  (like in
Fig.~\ref{dj1.fig}c)  mainly for large  scattering  angles where, with
the choice of $\mu =k(\chi)E_T/2$, the authors of  Ref.~\cite{Ellis92}
observe  that LO and NLO are quite  comparable.  This  underlines  the
assumption  given by Altarelli  {\em et al.}  \cite{Altarelli96}  that
the ratio $Z'/{\rm  QCD}$ should merely remain  unchanged (up to a few
percent) in a transition to NLO.

To  emphasise  the  influence of the $Z'$ even more, we  increased  in
Fig.~\ref{dj2.fig}  the dijet  masses  up to the  region  of  $M_{Z'}$
itself.  For $M_{jj}=1100$~GeV  (Fig.~\ref{dj2.fig}b) we calculate for
the dijet cross section in LO QCD:  $dN/(Nd\chi)|_{\rm  QCD} = 0.0363$
for   $\chi=1.5$   ($\Theta^\star=78^{\rm   o}$).  The   LO   QCD+$Z'$
calculation, however, yields a value of $dN/(Nd\chi)|_{{\rm QCD}+Z'} =
0.0610$,  which  means  an  increase  by a  factor  of 1.7 due to $Z'$
exchange.

It will be very  interesting  to  compare  our  predictions  to future
results  from the  Tevatron  to  decide  whether  the $Z'$  model is a
suitable   description  {\em  if}  an  excess  in  the  dijet  angular
distributions  for higher dijet masses  continues to be observed.  But
such an excess has to be expected after the single inclusive jet cross
section  measurements.  Such a double check would of course  underline
the  reliability  of  the  experimental  data  as  well  as  test  the
theoretical predictions by any other models.  We would like to mention
some   still   preliminary   data  taken  by  the  CDF   Collaboration
\cite{Chao96}.  The data are still  limited to dijet  masses for which
the $Z'$  contribution is not  significantly  standing out against the
statistical  and  systematic   errors,  even  though   especially  the
statistical  errors  could  be  quantitatively   further  reduced.  An
analysis of these  data\footnote{ I am indebted to C.~Wei from the CDF
Collaboration for providing me with these preliminary results.}, which
is nor presented  here, showed again the  excellent  agreement  with a
calculation in LO in  combination  with the  renormalisation  scale of
Eq.~(\ref{ellis}).

The ratios  $Z'/{\rm QCD}$ of our  calculations  are also presented in
Fig.~\ref{dj2.fig}.  This gives even  stronger  evidence  for the fact
that for higher dijet masses the $Z'$ contribution  especially governs
the larger  scattering  angles  whereas  for small  angles  the ratios
behave  smoothly.  This can be observed in  Fig.~\ref{dj2.fig}b  where
$|Z'/{\rm  QCD}|$ even  shrinks for larger  $\chi$ such that one might
conclude that for high dijet masses but very small  scattering  angles
the  $Z'$  contribution   becomes   irrelevant.  Even  though  the  LO
calculations are not quite  compatible to NLO in the high $\chi$ range
\cite{Ellis92},  the  corrections  due to NLO are  supposed to cancel,
considering  the ratios only, such that this  observation  should also
hold in a NLO calculation.

We conclude this section with a comparison to recent very precise data
from the D0  Collaboration  \cite{D094}.  In the  measured  dijet mass
range  $175$~GeV$<M_{jj}<350$~GeV  the effect of the $Z'$ is of course
negligible  as we have  learned  from the CDF data.  However,  as this
data are the most  precise  available at this stage, we might test our
argumentation  about the  reliability of the LO  calculations.  It has
been reported  \cite{D094} that the data are significantly  consistent
with NLO QCD  calculations.  In  Fig.~\ref{dj3.fig}  we present the D0
data and  normalise  our cross  sections as before in the shown $\chi$
range.  We  restrict  ourselves  to a  presentation  of  the  QCD+$Z'$
results only, as the  differences to pure QCD are not striking in this
mass regime (cf.  Fig.~\ref{dj1.fig}a).  The  numerical  values of the
calculation with  $\mu=k(\chi)E_T/2$ lie almost within the error bars.
Recall  that  this  choice  of  $\mu$ is in good  agreement  with  NLO
according   to   \cite{Ellis92}.   A   statistical   analysis   yields
$\chi^2=12.39$,  so the LO  calculation  satisfactorily  describes the
experimental  data,  exactly  as  has  been  claimed  throughout  this
section.  A picture of  consistency  emerges out of the  comparison to
the  experimental  data.  The dashed  line  shows the  result  for the
calculation  with  $\mu=E_T/2$.  The   similarity  in  $\chi^2$  is an
indicator of how reliably this scale is again working in approximating
NLO results for large scattering angles.

For  illustrative  reasons we also present the result for a completely
different  renormalisation  scale.  This shows  that a less  dynamical
scale like $\mu=M_{jj}$ cannot describe the experimental  results (the
$\chi^2$ value is also  presented).  The curve is nearly flat over the
whole $\chi$ range.


\section{The $Z'$ at the LHC}
\label{sec4.ms}

\begin{figure}[b]
\unitlength1cm 
\begin{center} 
\begin{picture}(13.9,16)
\makebox[13.2cm]{
\epsfxsize=13.2cm  
\epsfysize=11cm  
\epsffile{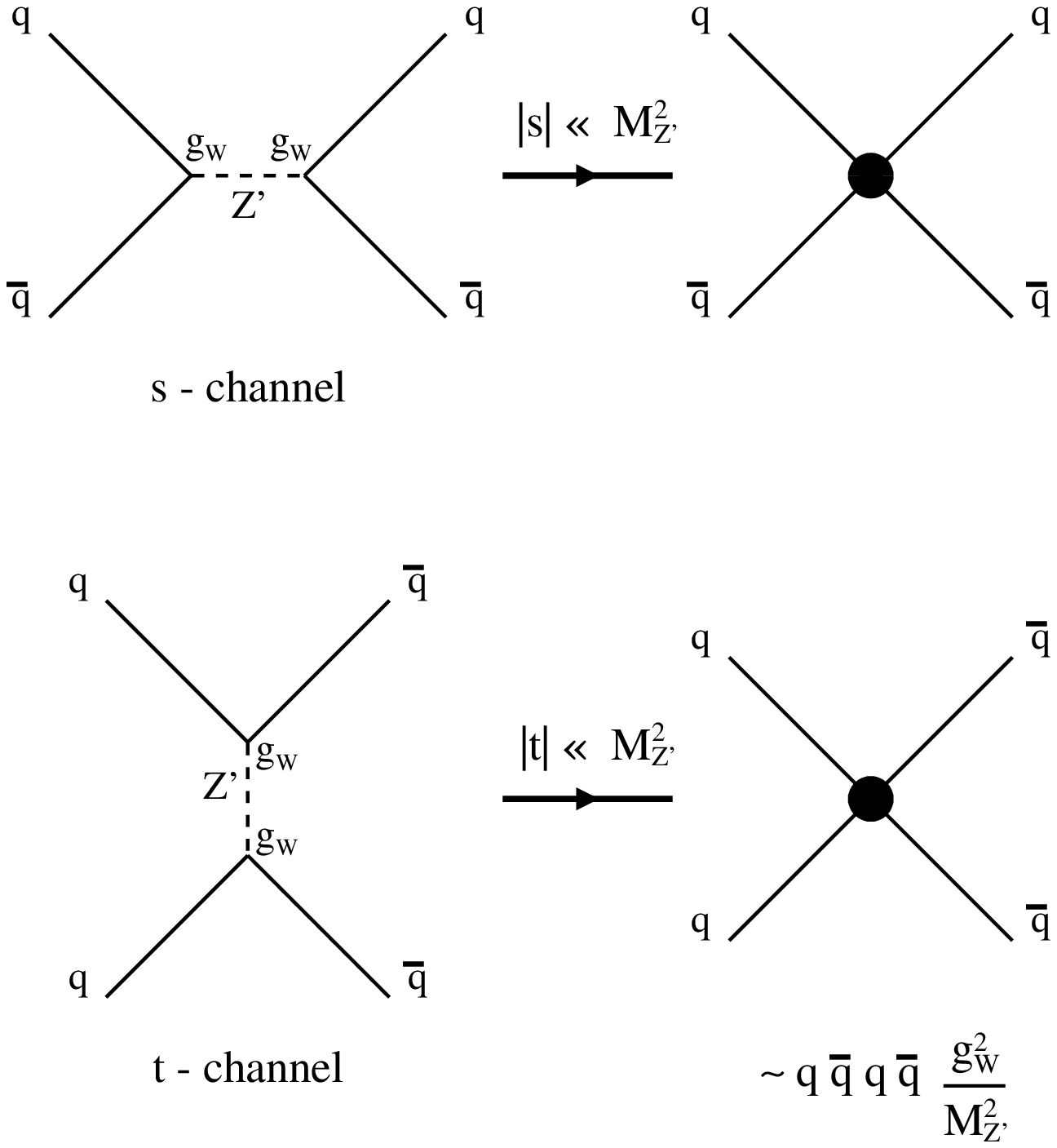}  }
\end{picture} 
\end{center}

\caption[]{\label{kin.fig}The  $s$--  and  $t$--channel  contributions
according  to $Z'$  exchange  (left  side).  For $|s|$ and $|t|$ being
small,  the $Z'$  acts  like an  effective  contact  interaction  with
relative strength $\sim \alpha_W/M^2_{Z'}$ (right side).}

\end{figure}

\begin{figure}[b] 
\unitlength1cm 
\begin{center} 
\begin{picture}(13.9,24)
\makebox[13.9cm]{
\epsfxsize=13.9cm  
\epsfysize=14cm  
\epsffile{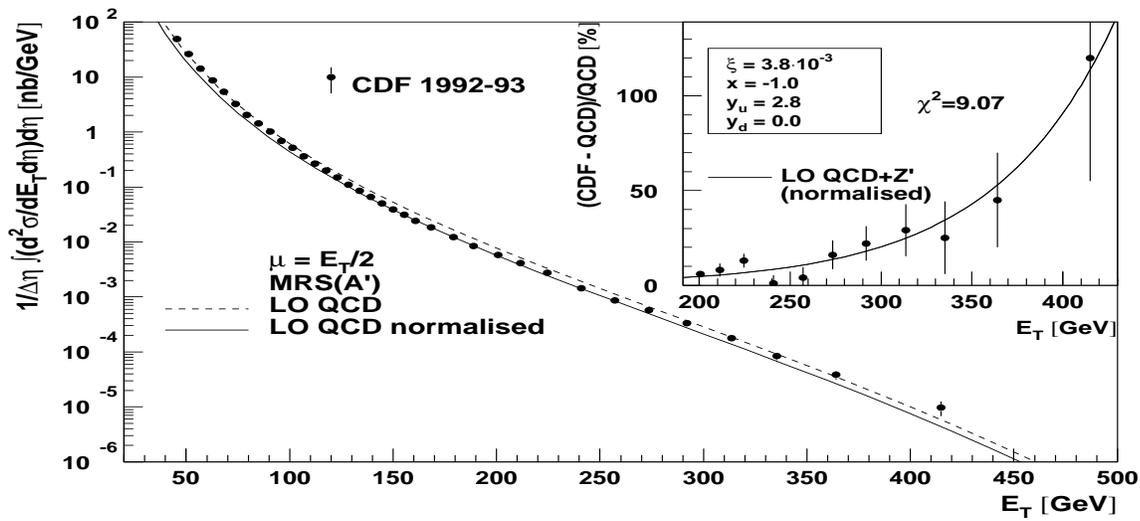}  }
\end{picture} 
\end{center}

\vskip -8cm

\caption[]{\label{lot.fig}LO  calculation of the single  inclusive jet
cross section  (dashed line) and the normalised LO fit (solid line) to
the CDF 1992--93 data  \cite{CDF93}  (as  discussed in the text).  The
small inset shows the  difference in per cent between our  calculation
and the measured cross sections by the CDF  Collaboration.  Also shown
is the {\em  best--fit} of the included $Z'$ model with the parameters
also presented (cf.  Section~\ref{sec3.1.ms}).  }

\end{figure}

\begin{figure}[b] 
\unitlength1cm 
\begin{center} 
\begin{picture}(13.9,16)
\makebox[13.9cm]{
\epsfxsize=13.9cm  
\epsfysize=14cm  
\epsffile{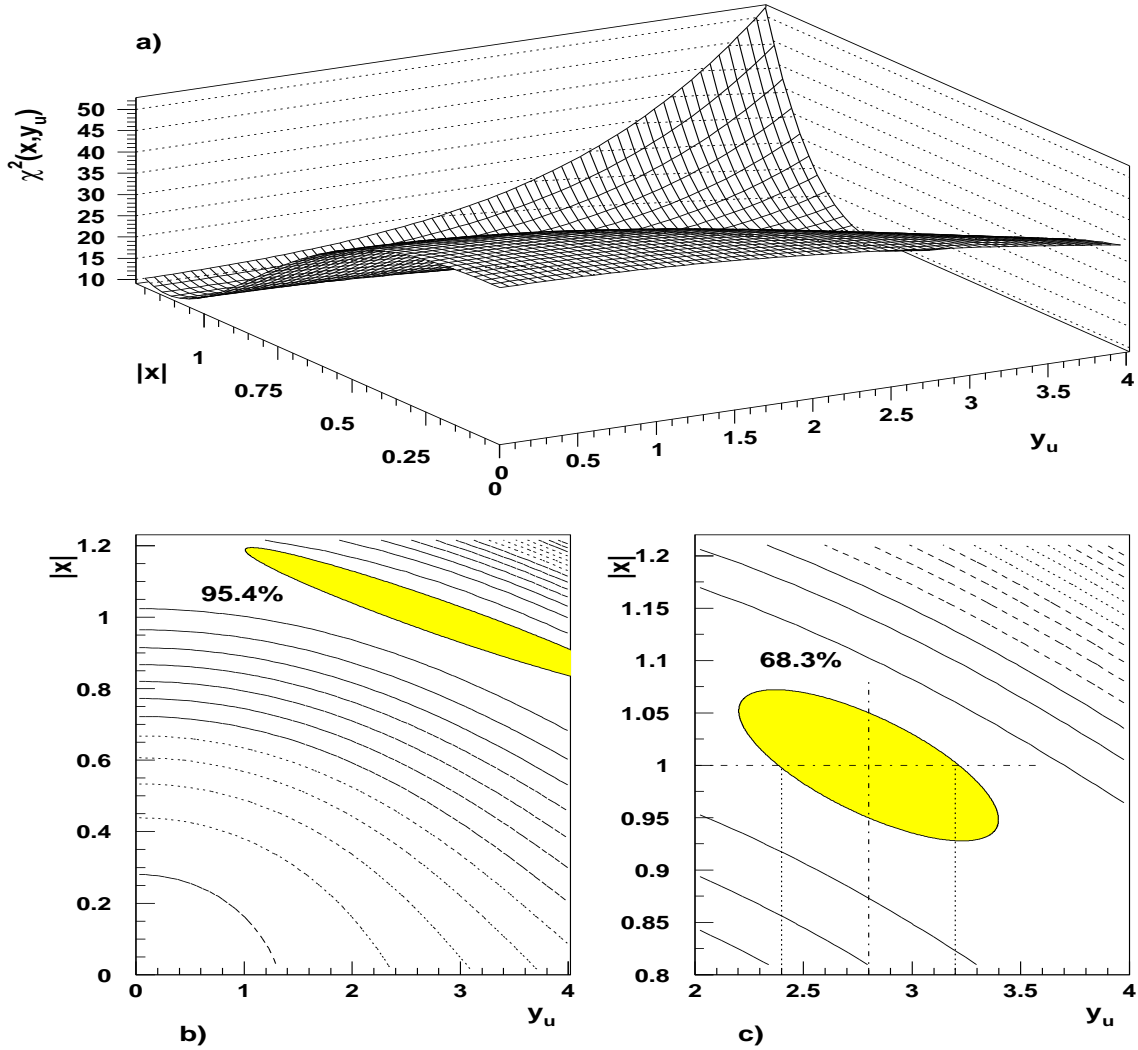}  }
\end{picture} 
\end{center}

\caption[]{\label{chi.fig}The   statistical   results   of  our   $Z'$
analysis:  (a) the  $\chi^2$  distribution  as a  function  of the two
degrees  of freedom  $x$ and $y_u$ (the  fitted  parameters),  (b) the
95.4\% confidence  ellipse and (c) the 68.3\% confidence  ellipse with
the central values $x=-1.0$ and $y_u=2.8$ being  indicated  (best--fit
values).}

\end{figure}

\begin{figure}[b] 
\unitlength1cm 
\begin{center} 
\begin{picture}(13.9,16)
\makebox[13.9cm]{
\epsfxsize=13.9cm  
\epsfysize=14cm  
\epsffile{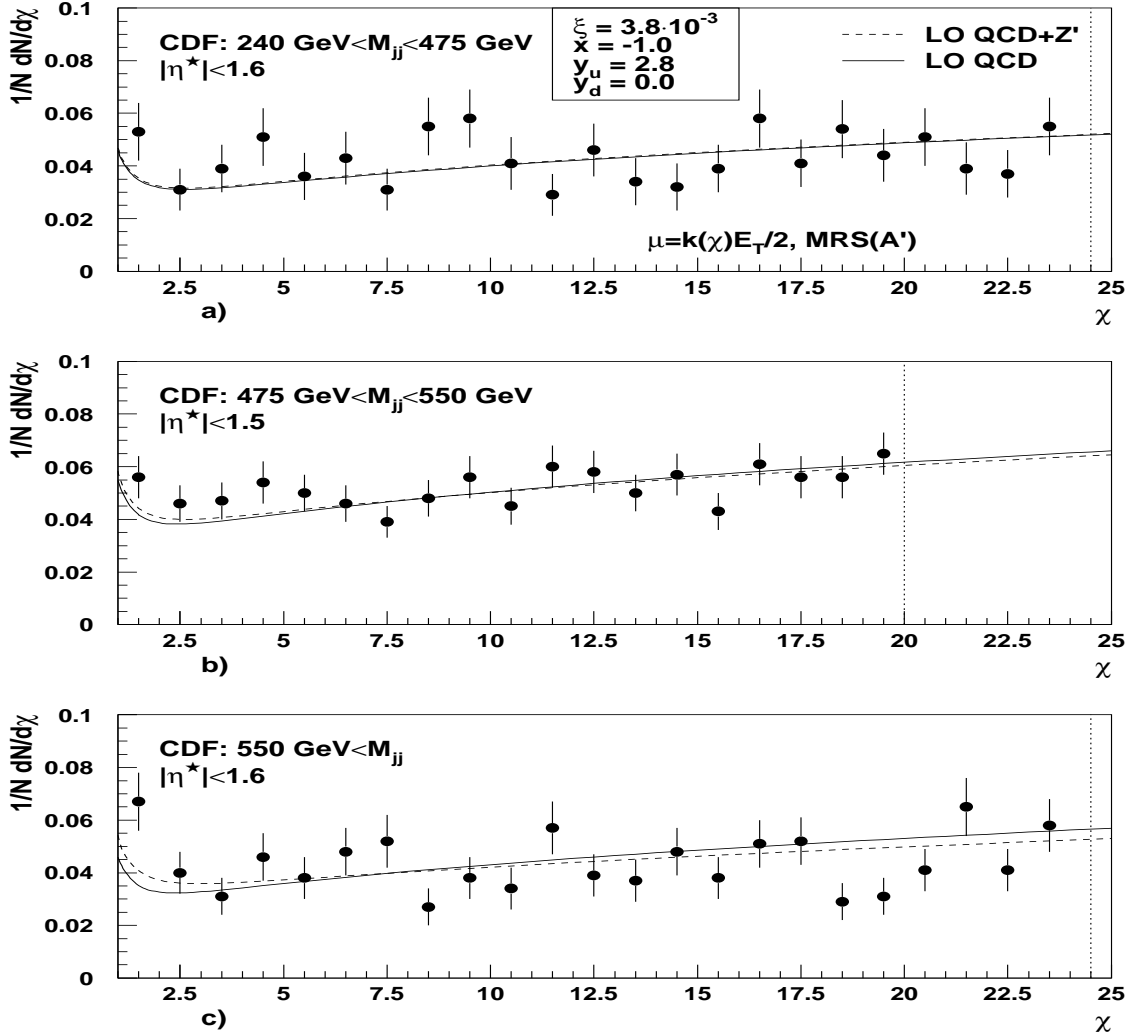}  }
\end{picture} 
\end{center}

\caption[]{\label{dj1.fig}The   normalised  dijet  cross  sections  at
${\cal{O}}   (\alpha_S^2)$   for  pure  QCD  (solid   lines)  and  the
additionally  coupled  vector  boson  $Z'$  (dashed  lines)  in  three
different  dijet  mass  bins:  (a)   $240$~GeV$<M_{jj}<475$~GeV,   (b)
$475$~GeV$<M_{jj}<550$~GeV  and (c)  $M_{jj}>550$~GeV.  The  numerical
results are  compared to the CDF~'92  measurements  \cite{CDF92}.  The
kinematical   constraints  on  $\eta^{\star}$  and  the  normalisation
intervals in $\chi$ are indicated and discussed in the text.  All $Z'$
calculations  were performed for the central  parameter fit:  $x=-1.0$
and   $y_u=2.8$.   As    renormalisation    scale   we   have   chosen
$\mu=k(\chi)E_T/2$ from Ref.~\cite{Ellis92}.}

\end{figure}

\begin{figure}[b] 
\unitlength1cm 
\begin{center} 
\begin{picture}(13.9,16)
\makebox[13.9cm]{
\epsfxsize=13.9cm  
\epsfysize=14cm  
\epsffile{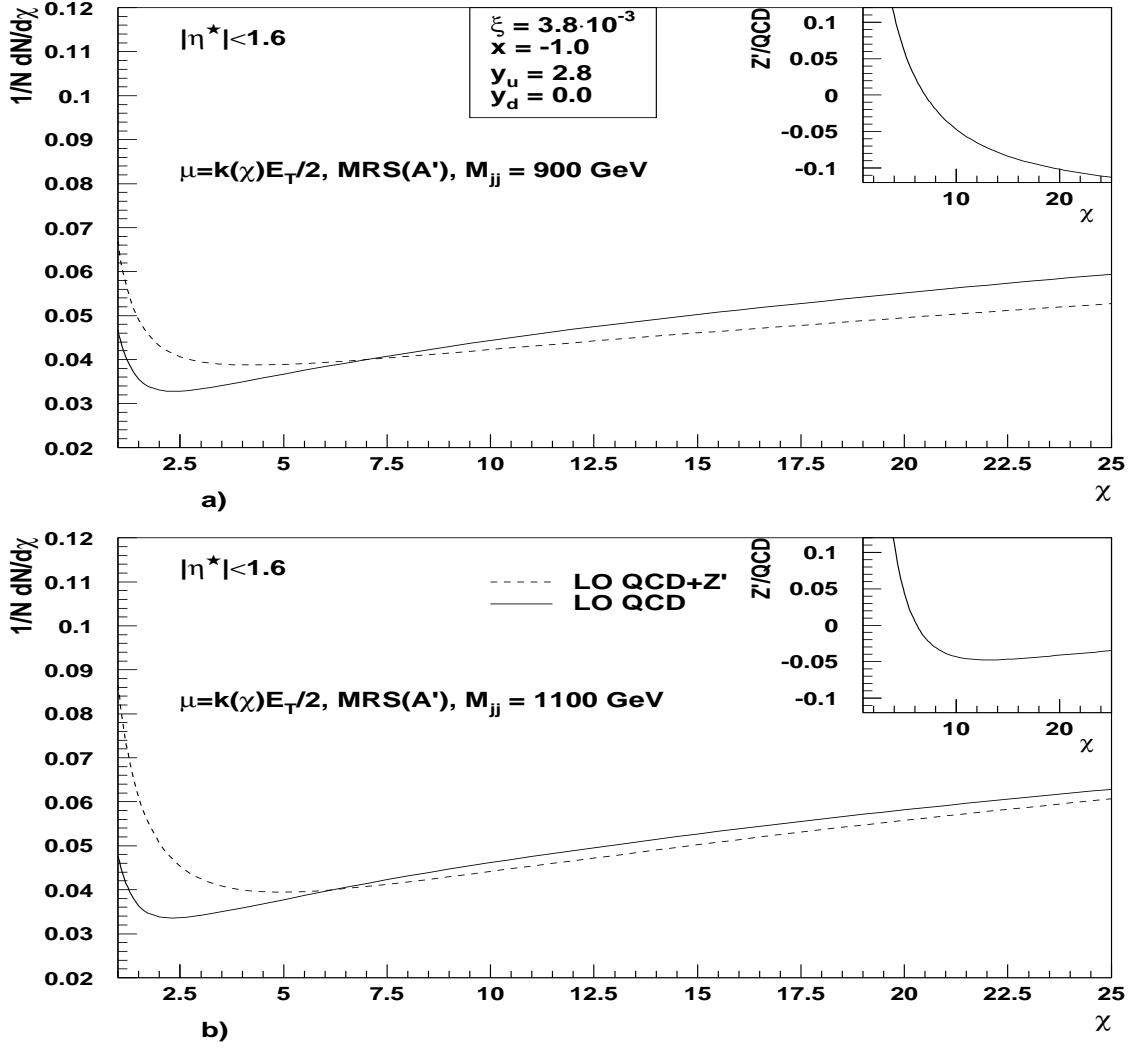}  }
\end{picture} 
\end{center}

\caption[]{\label{dj2.fig}Same  as Fig.~\ref{dj1.fig}, but now for the
two   fixed   dijet   mass   bins:  (a)   $M_{jj}=900$~GeV   and   (b)
$M_{jj}=1100$~GeV.  The dijet cross  sections are  normalised to unity
in the interval  $1\leq\chi\leq24.5$.  The relative  contributions  of
the  $Z'$  to  the  LO QCD  calculations  ($Z'/{\rm  QCD}$)  are  also
presented.}

\end{figure}

\begin{figure}[b] 
\unitlength1cm 
\begin{center} 
\begin{picture}(13.9,16)
\makebox[13.9cm]{
\epsfxsize=13.9cm  
\epsfysize=14cm  
\epsffile{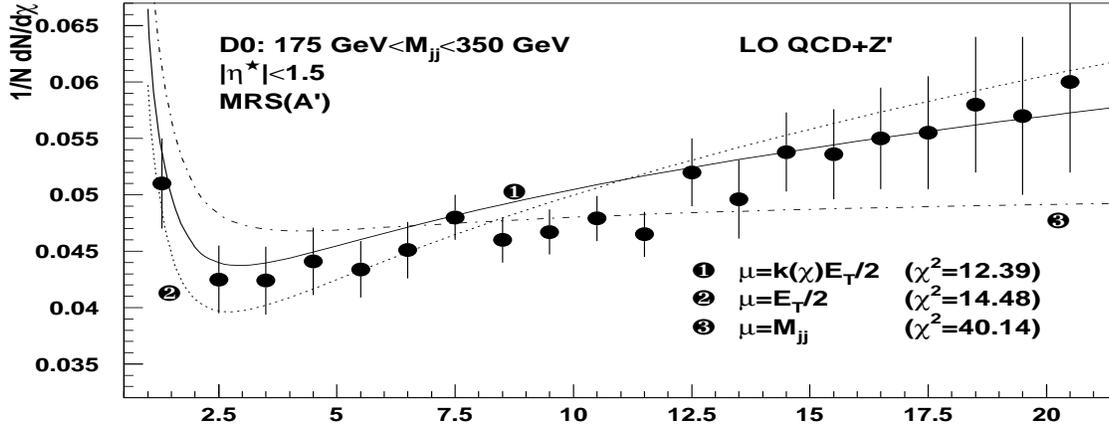}  }
\end{picture} 
\end{center}

\vskip -5.5cm

\caption[]{\label{dj3.fig}The  dijet angular  distributions in leading
order with three different  renormalisation scales including the scale
defined in  Eq.~(\ref{ellis})  and employed  throughout  this section.
The  cross  sections  are  integrated   over  $M_{jj}$  in  the  range
$175$~GeV$<M_{jj}<350$~GeV.  As before we present the {\em normalised}
cross  sections  but now for the LO  QCD+$Z'$  calculation  only.  The
results   are   compared   to  the   data   taken   from  the  D0  '94
measurements~\cite{D094}.}

\end{figure}

The  question we want to address in this  section is how the $Z'$ will
influence  the  measured  jet  cross  sections  at the  LHC.  From our
results  of  Section~\ref{sec3.ms}  we  expect  the  influence  to  be
generally  enhanced  due  to  a  higher  centre--of--mass   energy  of
$\sqrt{s}  =  10$--$14$~TeV.  This  allows the  observation  of higher
transverse  energies  $E_T$ and dijet  masses  $M_{jj}$.  On the other
hand we expect the background  contributions like Drell--Yan processes
\cite{Drell70},    production    of   mini--jets    \cite{Ragazzon96},
diffraction\cite{UA8},   etc.  to   become   larger   such   that  the
signal/background  ratio for the $Z'$  will be even more  reduced.  We
constructed  the $Z'$ such that it does not  couple  to  leptons,  and
Drell--Yan processes via $Z'$ exchange have to be completely excluded.
Another feature somehow obstructs the detectability of the $Z'$ at the
LHC:  at a $pp$--collider and high centre--of--mass  energies the main
contributions  to the  two--parton  jet events come from  subprocesses
involving gluons, like $gg\rightarrow gg(q\bar{q})$ and $gq\rightarrow
gq$.  But the $Z'$ does not couple to gluons.  And as antiquarks  only
appear as sea  quarks in the  proton we expect  the main  contribution
from the $Z'$ at the LHC to come from the  $t$--channel  exchange (cf.
Fig.~\ref{kin.fig}).

In the following we shall  perform {\em all}  calculations  in pure LO
for $\mu=E_T/2$ and expect the arguments of  Section~\ref{sec3.ms}  to
be still valid,  namely a difference  between LO and NLO for large jet
energies  by a constant  factor  only and an even  better  coincidence
between  LO and NLO in the case of dijet  production.  The latter  has
been checked numerically by employing again the renormalisation  scale
of   Eq.~(\ref{ellis})    and   the   previous   results   stated   in
Ref.~\cite{Ellis92}.  At least  for the  ratios  $({\rm  QCD}+Z')/{\rm
QCD}$ we do not expect evident  differences to NLO, as NLO corrections
are expected to cancel.

\begin{figure}[b] 
\unitlength1cm 
\begin{center} 
\begin{picture}(13.9,16)
\makebox[13.9cm]{
\epsfxsize=13.9cm  
\epsfysize=14cm  
\epsffile{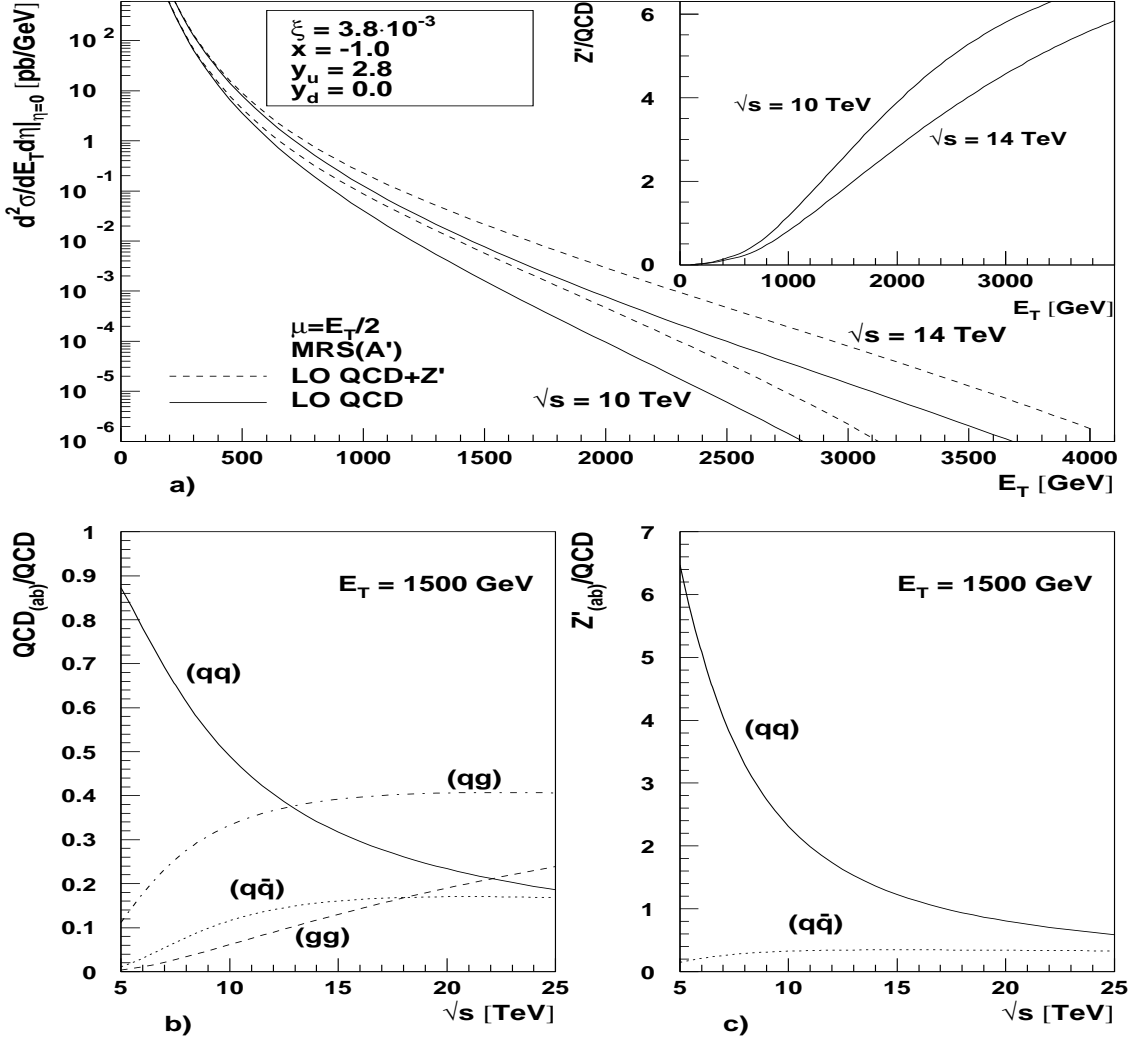}  }
\end{picture} 
\end{center}

\caption[]{\label{lhc1.fig}LO  calculation of the single inclusive jet
cross sections at the LHC for the two centre--of--mass energies 10~TeV
and 14~TeV.  The ratios  $Z'/{\rm QCD}$ are shown in the inset of (a).
The contributions of the different subprocesses $ab\rightarrow cd$ for
(b) LO QCD and (c) $Z'$, normalised to the full LO QCD calculation, as
a function of the  centre--of--mass  energy are shown.  The transverse
jet energy was fixed to be $E_T = 1500$~GeV.  }

\end{figure}

In   Fig.~\ref{lhc1.fig}a  we  present  the  results  for  the  single
inclusive  cross  section  at the LHC for fixed  $\eta=0$.  The  inset
shows the ratios  $Z'/{\rm  QCD}$ for two  different  centre--of--mass
energies   as   a   function   of   $E_T$.   We   observe   that   for
$E_T\sim1000$~GeV  the contribution  from the $Z'$ matches the QCD one
for both  curves.  The curves are then  rising  very  steeply  but the
typical  $\propto  E_T^4$  behaviour  we  observed  in  the  inset  of
Fig.~\ref{lot.fig}    for   the    Tevatron    is    suppressed    for
$E_T\stackrel{>}{\sim}2500$~GeV.   To   understand   the    underlying
mechanism for this observation we present in Figs.~7b,c the individual
subprocesses  $ab\rightarrow cd$ for the QCD and the $Z'$ calculation.
For higher centre--of--mass  energies the gluons play the pivotal role
and dominate the matrix elements of Eq.~(\ref{single}).

At typical LHC energies the $qg\rightarrow qg$ contribution  dominates
with  about  40\% of all other  subprocess  events.  For still  larger
values of  $\sqrt{s}$  also the  gluon--gluon  fusion rate is linearly
growing  whereas  the  number  of  subprocesses  including  quarks  or
antiquarks   as   initial   partons   is   diminished   as   shown  in
Fig.~\ref{lhc1.fig}b.  We also observe the ratio $(q\bar{q})/(qg)=4/9$
as    predicted    by    perturbative    QCD    \cite{Brodsky76}    in
Fig.~\ref{lhc1.fig}b.

The $Z'$ does not couple to gluons and therefore the $Z'$ contribution
is rising  more  slowly for higher  centre--of--mass  energies  as the
gluons  actually give the dominant  contributions.  The  corresponding
subprocesses   governing   the  $Z'$   contribution   are   shown   in
Fig.~\ref{lhc1.fig}c.  This  explains  two  features   observable   in
Fig.~\ref{lhc1.fig}a:  first, the ratio  $Z'$/QCD is becoming  flatter
for higher values of $\sqrt{s}$  and second, the main high  transverse
jet energy is carried by the  gluons.  The latter is a well known fact
and  was  theoretically  dealt  with  in  Ref.~\cite{Antoniou86}.  The
relative  contributions  of  quarks  and  antiquarks  to  large  $E_T$
processes is small, which yields the observed  smoothing in the ratios
at  larger  $E_T$.  Note  the  absolute   scales  in  Figs.~7b,c.  For
$\sqrt{s}=10$~TeV the $Z'_{(qq)}$ subprocess exceeds the corresponding
${\rm  QCD}_{(qq)}$  rate by a  factor  of five.  Fig.~\ref{lhc1.fig}c
also demonstrates the predominance of the $Z'$  $t$--channel  exchange
compared to the $s$--channel exchange sketched in  Fig.~\ref{kin.fig}.

\begin{figure}[b] 
\unitlength1cm 
\begin{center} 
\begin{picture}(13.9,16)
\makebox[13.9cm]{
\epsfxsize=13.9cm  
\epsfysize=14cm  
\epsffile{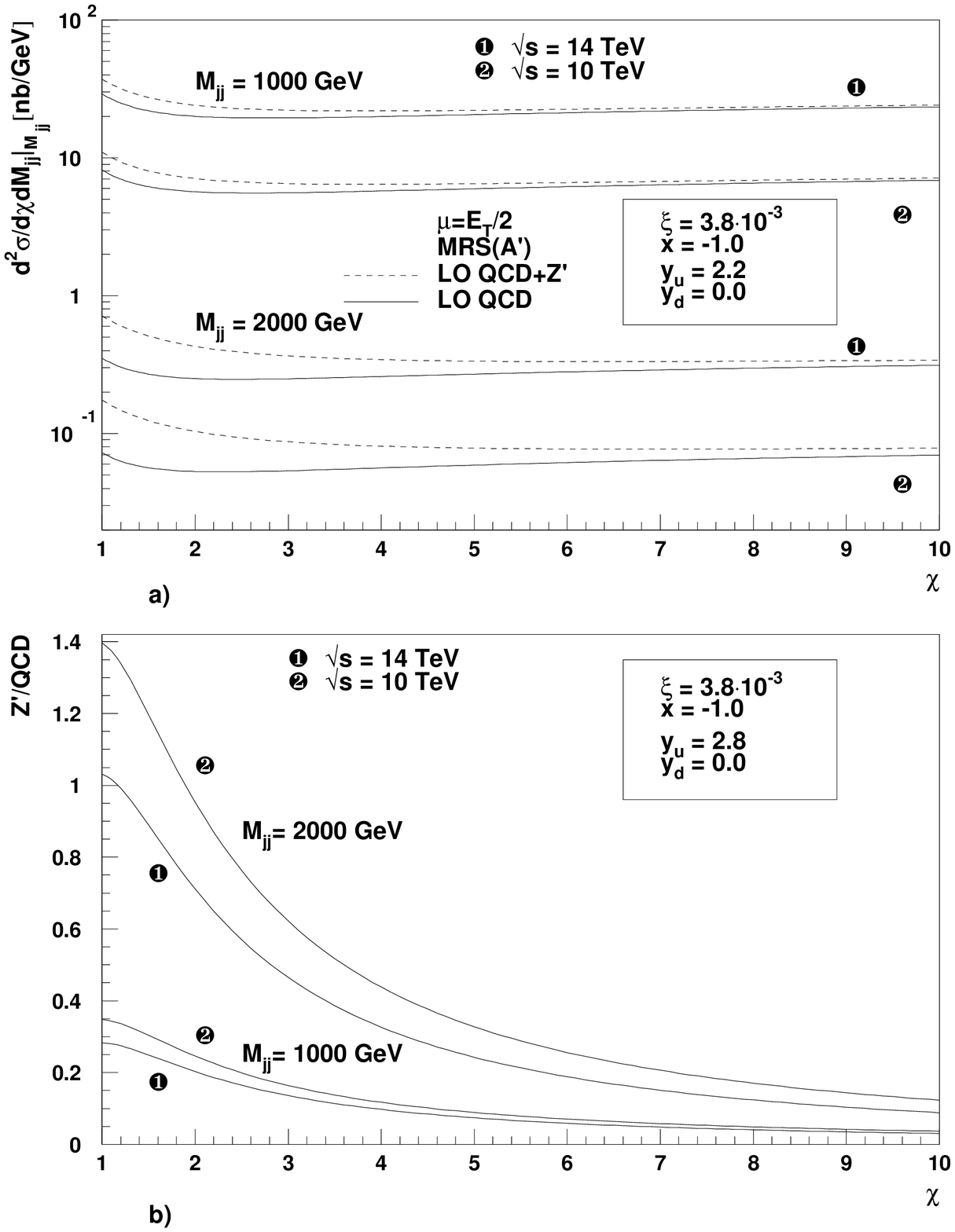}  }
\end{picture} 
\end{center}

\caption[]{\label{lhc2.fig}The  dijet angular distributions at the LHC
for two  different  invariant  dijet masses.  The  unnormalised  cross
sections  are shown in (a) for LO QCD and LO QCD+$Z'$.  In (b) we show
the corresponding ratios $Z'$/QCD again for the central fit parameters
of the $Z'$ model.}

\end{figure}

We also give predictions for the dijet angular distributions as we did
for  the  Tevatron.  Fig.~\ref{lhc2.fig}a  shows  the  results  for  a
calculation with $M_{jj}=1000$~GeV and $M_{jj}=2000$~GeV again for the
two different centre--of--mass energies.  Unlike the presentations for
the Tevatron we now show the {\em unnormalised}  distributions for our
best--fit  parameters  (\ref{parameters}).  Qualitatively  we find the
same  results  as for  the  Tevatron:  the  $Z'$  boson most  strongly
influences the small $\chi$ region (again we interpret the $Z'$ acting
as an {\em effective  contact  interaction}  \cite{Eichten89}  in this
regime (cf.  Fig.~\ref{kin.fig}),  contracting  its  propagator  to an
effective  four--fermion  point--like  interaction) and this effect is
again  enhanced for higher  dijet  masses.  The  corresponding  ratios
shown in Fig.~\ref{lhc2.fig}b  underline the conclusions already drawn
for the Tevatron, but now on a much larger scale.

\begin{figure}[b] 
\unitlength1cm 
\begin{center} 
\begin{picture}(13.9,20)
\makebox[13.9cm]{
\epsfxsize=13.9cm  
\epsfysize=14cm  
\epsffile{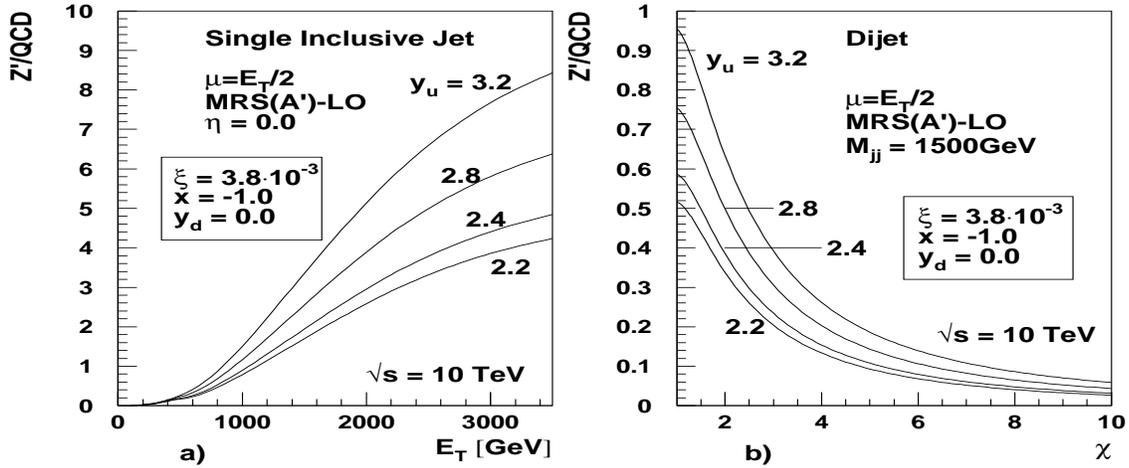}  }
\end{picture} 
\end{center}

\vskip -4cm

\caption[]{\label{lhc3.fig}The  ratios  $Z'$/QCD  for the  (a)  single
inclusive  jet  cross  sections   ($\eta=0$)  and  (b)  dijet  angular
distributions  ($M_{jj}=1500$~GeV)  at the LHC.  We keep $x, y_d, \xi$
and  $M_{Z'}$  fixed to the  values of our  best--fit  and vary  $y_u$
according    to   the    68.3\%    confidence    ellipse    shown   in
Fig.~\ref{chi.fig}c.  We  also  present  the   calculations   for  the
best--fit     value     $y_u=2.2$     of     Altarelli     {\em     et
al.}~\cite{Altarelli96}.}

\end{figure}

Because  we have  so far  presented  our  numerical  results  for  our
best--fit values  (\ref{parameters}) only, we finally want to show the
variations  of the $Z'$ impact due to upper and lower bounds in accord
with our  analysis.  If we fix  $x=-1.0$,  as we found the central $x$
value to be, then we get upper  and  lower  bounds  on $y_u$  from our
$\chi^2$  analysis if we restrict  our  fit--acceptance  to the 68.3\%
confidence ellipse shown in Fig.~\ref{chi.fig}c.  For $x=-1.0$ we read
off  $y_u\in\lbrack  2.4,3.2\rbrack$.  Fig.~\ref{lhc3.fig}a  shows the
single  inclusive  jet  ratios  for  the  three  different  values  of
$y_u=2.4,  2.8$ and $3.2$  being the lower  bound,  central  value and
upper  bound  respectively.  The  discrepancy  between  the  different
choices of $y_u$ becomes very  striking for higher $E_T$  values.  The
total decay width varies from  $\Gamma_{Z'}=508.0$~GeV  ($y_u=2.4$) up
to  $\Gamma_{Z'}=801.4$~GeV  ($y_u=3.2$),  which  increases  the phase
space of the $Z'$ especially at high  transverse  energies.  So, large
$E_T$  measurements  at the LHC  might be an  excellent  probe to more
precisely  fix the value of  $y_u$,  as the  cross  sections  are very
strongly  dependent on $y_u$ in this energy range and so a clear $y_u$
correspondence  is  achievable.  The  difference  to the  best--fit of
Altarelli {\em et al.}  \cite{Altarelli96}  ($y_u=2.2$) is also shown.
Note the  difference  of only 7\% to our lower bound  ($y_u=2.4$)  for
$E_T=3000$~GeV.

Fig.~\ref{lhc3.fig}b  finally shows the ratios  $Z'$/QCD for the dijet
angular   distributions   with  the  same   values  for  $y_u$  as  in
Fig.~\ref{lhc3.fig}a.  The $Z'$ impact on the small  $\chi$  region is
again  significant.  The extreme values of $y_u$ differ by a factor of
roughly  two  in  the  complete  $\chi$  range  shown.  Again,  future
measurements  of the  dijet  angular  distributions  at the LHC  might
further determine $y_u$ more exactly according to the large dependence
of the ratios to the choice of this coupling parameter.


\section{Conclusions}
\label{sec5.ms}

In   this    paper   we    exploited    the   idea    suggested     in
Refs.~\cite{Altarelli96,Chiapetta96}   to  give   predictions   for  a
postulated  new  heavy  vector  boson  $Z'$  at  the  LHC.  With  this
additional  very  massive  boson  it was  possible  to  quantitatively
explain the reported $R_{b,c}$ anomalies from LEP1/SLC  experiments as
well as simultaneously the measured CDF jet excess rate.  It was shown
by above  authors  that the  postulated  vector  boson must have three
special  features:  it is  leptophobic  and couples very strongly, but
family--independent,  to $u$-- and  $d$--type  quarks; it shows a weak
mixing with the standard $Z$ gauge boson in order to contribute to its
decay widths  $\Gamma(Z\rightarrow  b\bar{b},c\bar{c})$ in particular;
it is very massive with a typical mass of order $M_{Z'}=1$~TeV.

In this work we fitted the  coupling  parameters  $x$ and $y_u$ of the
$Z'$ in a global leading order  $\chi^2$  analysis to the 1992--93 CDF
data on the single  inclusive jet cross  sections.  Although we find a
slightly   larger  value  for  $y_u$  than   Altarelli  {\em  et  al.}
\cite{Altarelli96},  we showed that our best--fit parameters are still
in better  accordance  with the LEP1 $R_{b,c}$  measurements  than the
Standard Model predictions.

With  this set of  parameters  we then gave  predictions  for the $Z'$
effect on future  precision  measurements  at the LHC.  We showed  the
corresponding physical parameter ranges for which the influence of the
$Z'$ is expected to be most  striking  and besides  {\em  qualitative}
considerations  we also provided {\em  quantitative}  predictions  for
single inclusive jet cross sections and angular dijet distributions at
the  LHC.  We  presented  numerical  results  for  different  coupling
parameters $y_u$ that were allowed on the 68.3\% confidence level from
our previous  CDF data fit.  This will help to further  determine  the
free  parameters  of the $Z'$  model  as soon as first  LHC  data  are
available.

As a final critical  remark we want to point out that despite the very
precise and reliable  experiments  there might still be no  compelling
reason  to look  for {\em  new  physics}.  However,  future  data  are
necessary,  and the LHC will  play a  pivotal  role as a  high--energy
laboratory and new  theoretical  models and  predictions,  rising from
such  fundamental  contradictions  to the Standard  Model, will become
important.

We did not try to answer  the  question  of where  the  $Z'$, if it is
indeed   genuine,   originates   from.  For  an  overview  on  several
motivations  for the existence of additional  vector bosons and a list
of the most studied  models we refer to  \cite{Hewett89}.  In addition
we  should   mention  a  model  for  the  neutral  boson  proposed  in
\cite{Foot90},  where it  originates  from the breaking of an extended
colour group, such as ${\rm SU(4)}_{\rm C}$ or ${\rm  SU(5)}_{\rm C}$.
In this model the vector boson is very strongly  coupled to $q\bar{q}$
pairs and weakly  coupled to leptons.  As  reported in  \cite{Rizzo93}
its mass  should  be  larger  than  600~GeV.  In view of the  proposed
features this model could be a promising $Z'$ candidate.


\section*{Acknowledgements}

First of all I would  like to thank  James  Stirling  for  drawing  my
attention to this topic and for many fruitful discussions  throughout.
Chao Wei  from  the CDF  Collaboration  is  thanked  for  making  some
preliminary data on the dijet angular distributions  accessible to me,
to cross-check the stated  assumptions about LO  approximations.  I am
indebted  to John  Campbell  and  Colin  Weir  for  critical  comments
concerning the manuscript.  Finally,  financial support in the form of
a    ``DAAD--Doktorandenstipendium''    HSP--II/AUFE   is   gratefully
acknowledged.

\newpage



\newpage


\end{document}